\documentclass[aps,prl,twocolumn,groupedaddress]{revtex4}

\usepackage{amsthm}
\usepackage{amssymb}
\usepackage{amsxtra}     
\usepackage{amsmath}
\usepackage[pdftex]{graphicx}
\usepackage{epstopdf}

\DeclareGraphicsExtensions{.pdf, .jpg, .tif, .eps}

\begin{document}


\title{The Micromechanics of Three Dimensional Collagen-I Gels}


\author{Andrew M. Stein}
\affiliation{Institute for Mathematics and its Applications, University of Minnesota, Minneapolis, MN 55403, USA}
\email[]{astein@ima.umn.edu}
\homepage[]{www.ima.umn.edu/~astein}
\author{David A. Vader, David A. Weitz}
\affiliation{School of Engineering and Applied Sciences, Harvard University, Cambridge, MA 02138, USA}
\author{Leonard M. Sander}
\affiliation{Michigan Center for Theoretical Physics and Department of Physics, University of Michigan, Ann Arbor, Michigan 48109, USA}


\date{\today}
\begin{abstract}  
We study the micromechanics of collagen-I gel with the goal of bridging the gap between theory and experiment in the study of biopolymer networks.  Three-dimensional images of fluorescently labeled collagen are obtained by confocal microscopy and the network geometry is extracted using a 3d network skeletonization algorithm.    Each fiber is modeled as a worm-like-chain that resists stretching and bending, and each cross-link is modeled as torsional spring.  The stress-strain curves of networks at three different densities are compared to rheology measurements.  The model shows good agreement with experiment, confirming that strain stiffening of collagen can be explained entirely by geometric realignment of the network, as opposed to entropic stiffening of individual fibers.  The model also suggests that at small strains, cross-link deformation is the main contributer to network stiffness whereas at large strains,  fiber stretching dominates.  Since this modeling effort uses networks with realistic geometries, this analysis can ultimately serve as a tool for understanding how the mechanics of fibers and cross-links at the microscopic level produce the macroscopic properties of the network.  While the focus of this paper is on the mechanics of collagen, we demonstrate a framework that can be applied to many biopolymer networks.
\end{abstract}

\pacs{}

\maketitle

\begin{figure}
		\centering
		\includegraphics[width=3in]{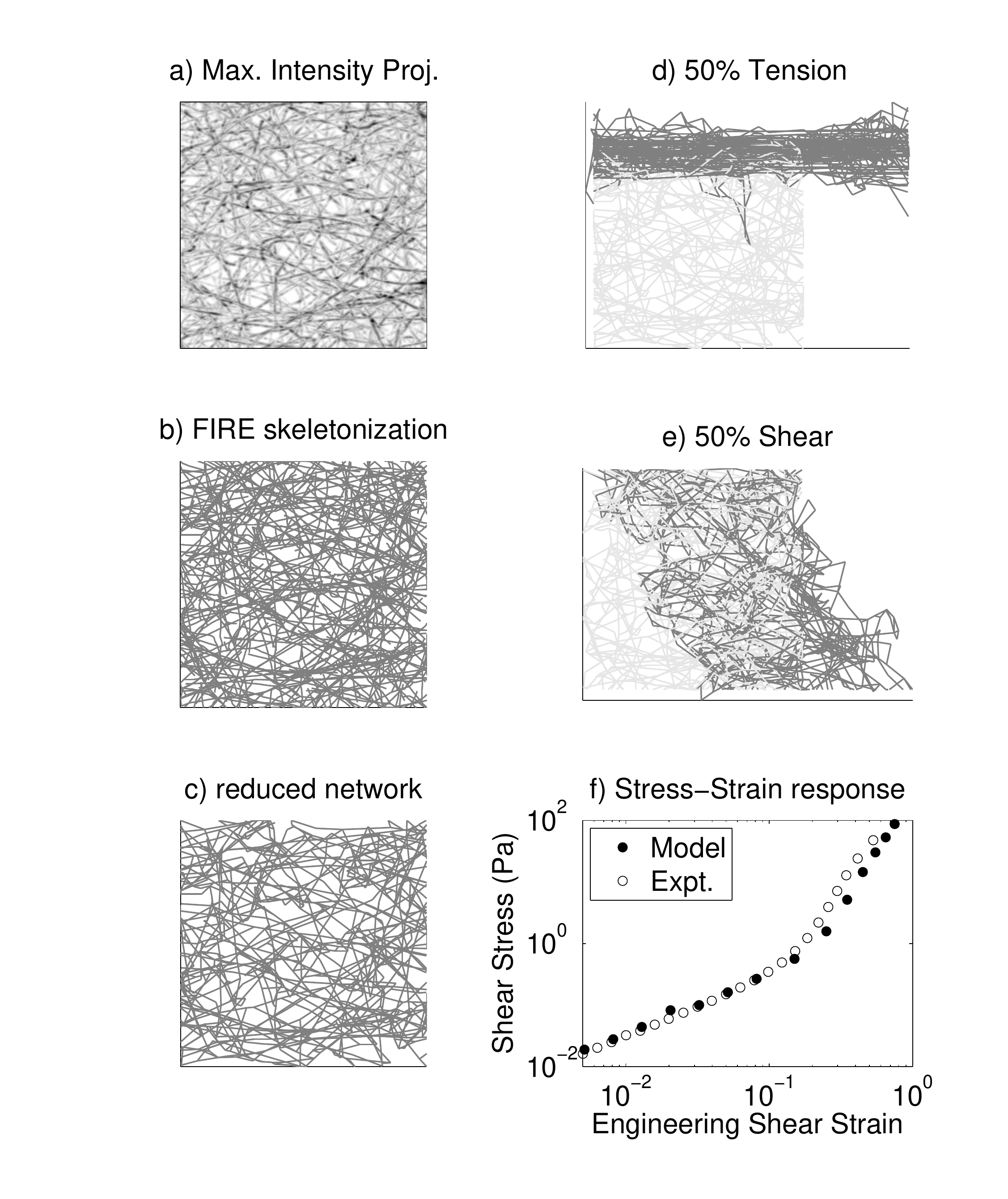}
		\caption{A typical gel (1.2 mg/ml, 25.6 $\mu$m $\times$ 25.6 $\mu$m $\times$ 25.6 $\mu$m).  a) Maximal Intensity Projection along the axis that is perpendicular to the focal plane of the microscope.  b) Projection of 3d network extracted by FIRE. c) Reduced network, where elements that don't contribute to the network stiffness have been removed for improved computational efficiency.  d) Deformation after 50\% tension.  e) Deformation after 50\% shear.  f)  Comparison of the stress-strain response between the model and experiment. \label{fig:summary}}
\end{figure}

Collagen is the most abundant animal protein \cite{Lodish99} 
and its mechanics have been studied in great detail \cite{Fung93}.   It takes on many morphologies, including skin, tendons, ligaments, individual fibers, and gels.  Of particular interest is the mechanics of collagen-I gels, shown in Figure~\ref{fig:summary}a.  These gels provide a relatively simple structure that can be noninvasively observed by confocal microscopy \cite{Stein08_fire} and used as a scaffold for growing artificial tissues \cite{Chandran04}, and as a 3d environment for studying cell motility \cite{Friedl98} and tumor invasion \cite{Kaufman05, Stein07_bj}. A critical first step in understanding these systems is to develop a model for the collagen gel alone.  In this paper we give a successful theoretical model of the micromechanics of realistic networks.

Collagen-I gels belong to a class of materials known as biopolymers.  Other examples include actin, found in the cytoskeleton and fibrin, a component of blood clots.  A common feature of biopolymer networks is their ability to strain stiffen by 2-3 orders of magnitude at large strains (Figure~\ref{fig:summary}f).  The cause of this strain stiffening is not well understood. 
Storm et al. \cite{Storm05} attributed strain stiffening in all biopolymer networks to the pulling out of entropic modes of individual filaments.  Their calculation required the assumption that deformations are affine.  Later, Heussinger et al. \cite{Heussinger06, Heussinger07} showed how one could deconstruct the network deformation into a set of floppy modes.  They concluded that accounting for the non-affinity was necessary in describing the elastic properties of the network.  Onck and colleagues \cite{Onck05, Huisman07} have proposed the alternative hypothesis that strain stiffening is due to the rearrangement of the fibers as the network is strained (Figures~\ref{fig:summary}d and \ref{fig:summary}e).  Resolving this debate has been difficult since almost all theoretical analysis has been on artificially generated networks.  The few examples of quantitative comparisons to experiment in the literature \cite{Storm05, Chandran06, Stylianopoulos07} are not able to quantitatively fit the full stress-strain response of the gel at varying densities using a single set of parameters.  

In this paper, we bridge the gap between model and experiment.  Three dimensional images of fluorescently labeled collagen gels at different densities are imaged by confocal microscopy (Figure~\ref{fig:summary}a) and the network geometry is extracted using a custom FIbeR Extraction (FIRE) algorithm (Figure~\ref{fig:summary}b) \cite{Stein08_fire}.   The gel is modeled as a random network of cross-linked fibers, as described below, and the stress-strain response is compared to that measured by an AR-G2 rheometer. 
Good agreement between model and experiment is obtained by fitting a single parameter, the cross-link stiffness.  The experiments are described in detail in \cite{Stein08_fire}.


In the model for the collagen gel, each fiber is treated as a discrete worm-like-chain (WLC) \cite{Klapper98} which resists stretching and bending, and each cross-link is treated as a torsional spring, thus more stiff than a freely rotating pin joint but less stiff than a welded joint of fixed angle.  The stretching modulus of an individual fiber is given by $K_s = EA$, where $E$ is the Young's modulus and $A$ is the cross-sectional area.  The Young's modulus of a fiber in aqueous conditions has been estimated to be between 30-800 MPa \cite{Graham06,Miyazaki99,vanderRijt06} and we use a modulus of 50 MPa, which fits the data well and is also close to the value chosen by Stylianopoulous et al.  (79MPa) to fit their model \cite{Stylianopoulos07}.  It has been shown that a single fiber will stiffen by a factor of 2-4 when strained \cite{vanderRijt06}.  We choose here to use a constant $E$ both to reduce the number of parameters in the model and to see if geometric reorientation of the network is enough to explain strain stiffening.  Stylianopoulos and Barocas \cite{Stylianopoulos07} also explored the bilinear and exponential constitutive relations for the individual fibers and observed only minor effects on the macroscopic network behavior. The radius of each fiber is $r = 30$ nm \cite{Stein08_fire, Raub07}.  The bending modulus of the fiber is given by $K_b = EI = 32$ pN-$\mu$m$^2$, where $I = \pi r^4 / 4$  \cite{Yang08}.  No cross-linking agent has been added to the gel and very little is known about the nature of the naturally formed collagen cross-links.  We find that we can fit all the data by setting the torsional spring stiffness to $K_x = 300 $ pN-$\mu$m.  To compare $K_b$ to $K_x$, we consider $K_b/l_c$, where the mean cross-link spacing is given by $l_c \sim 2 $ $\mu$m \cite{Stein08_fire}.  Thus, we find that $K_x \sim 20 K_b/l_c$.  One possible reason for a larger $K_x$ could be an increase in fiber radius near the cross-links by a factor of 2, since bending stiffness scales by $r^4$.


We assume that in the undeformed state of the network, there are no internal stresses.  Thus the fibers have an innate curvature and the cross-links have an equilibrium angle equal to that in their initial configuration.  We ignore entropic contributions to the fiber mechanics.  While the geometric persistence length of these fibers has been measured to be 20 $\mu$m \cite{Stein08_fire}, the thermal persistence length is much longer $l_p = K_b/kT \sim 1$ cm.  Furthermore, in the case that the strain stiffening is dominated by thermal compliance, one would expect to see a decrease in the yield strain with increasing concentration \cite{Head03pre}. 
Collagen gels, however,  have been shown to have a constant yield strain of about 60\% for a wide range of concentrations \cite{Roeder02}.  Thus the total energy in the network for a given configuration is given below.  
\begin{equation}
	U  	
	= \sum_{i=1}^{N_s} \frac{K_s}{L^i} \frac{\left(\Delta L^i\right)^2}{2}
	+ \sum_{i=1}^{N_b} \frac{K_b}{L^i} \frac{\left(\Delta \theta^i_b \right)^2}{2}
	+\sum_{k=1}^{N_x} K_x \frac{\left( \Delta \theta^i_x \right)^2}{2}
	\label{eq:U}
\end{equation}
Here $N_a$ is the number of elements of type $a \in \{s,b,x\}$, which denotes stretching, bending, and cross-link, $L_i$ is the length of stretching element,  $\theta^i_b$ and $\theta^i_x$ are the bending and cross-link angles respectively, and $\Delta$ indicates the difference between the deformed and undeformed state. 

To calculate the stress-strain relationship of our model network, we perform a series of 18 incremental strain steps by imposing a small deformation on one face, $F_1$ while holding the opposite face, $F_0$ fixed.  We impose two types of deformations: tension (Figure~\ref{fig:summary}d) and shear (Figure~\ref{fig:summary}e).  In a tensile deformation, we allow the vertices on $F_0$ and $F_1$ to move freely in directions perpendicular to the imposed strain to allow for perpendicular contraction that is seen to occur in these experiments \cite{Roeder02, Krishnan04}.  In experiments of this type, the distance between $F_0$ and $F_1$ is on the order of centimeters and the simulated network represents a small region near the center of a sample.  In shear, we do not allow the boundary nodes on $F_0$ and $F_1$ to move freely.  We compare the shear results to cone-plate rheometer experiments, where the shear faces are bound to the rheometer.   Here, the distance between $F_0$ and $F_1$ is 109 $\mu$m, and the simulated network is one fourth the length of the experimental sample between the boundaries.  In both deformations, all other nodes, including those on the four remaining faces of the network, are free to move. 

\begin{figure}
		\centering
		\includegraphics[width=2in]{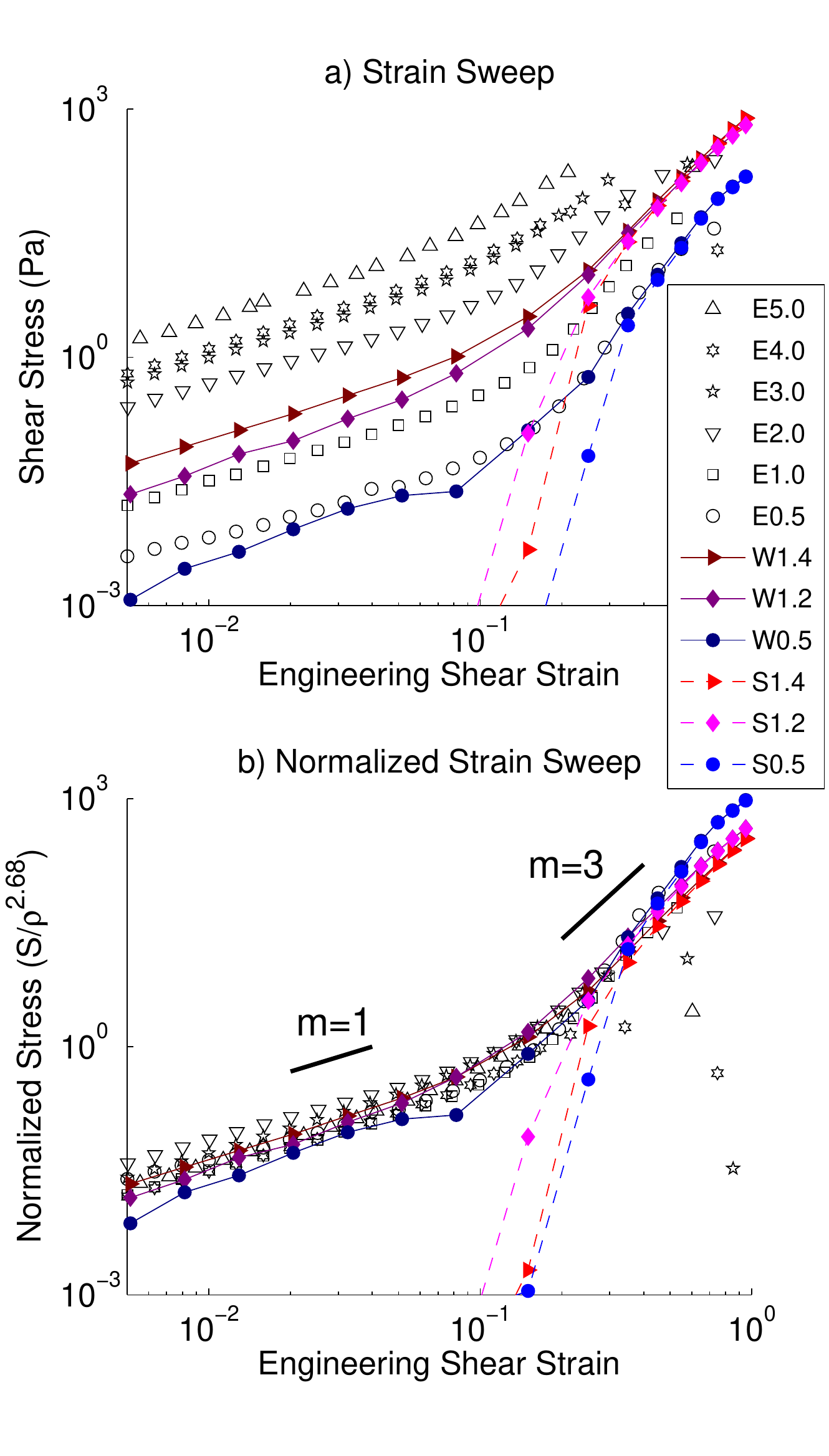}
		\caption{Stress-strain sweeps of the gel, from 0.5\% to 100\% strain.  In the legend, E denotes experiment, W denotes the WLC model, S denotes the spring model where we set $K_x = K_b = 0$, and the number denotes the collagen density.  a) Unscaled results.  Note the good agreement between model and data at small and large strains.  b) When the curves are scaled by $\rho^{2.68}$ relatively good data collapse is achieved.  We denote lines of slope $m=1$ and $m=3$ to guide the eye.  At large strains, scaling breaks down and the low density curves overtake the high density curves because at large strains, stiffness scales linearly with density.  Thus this rescaling serves mainly as a visualization tool and does not represent a true data collapse. \label{fig:strainsweep}}
\end{figure}

\begin{figure}
		\centering
		\includegraphics[width=3in]{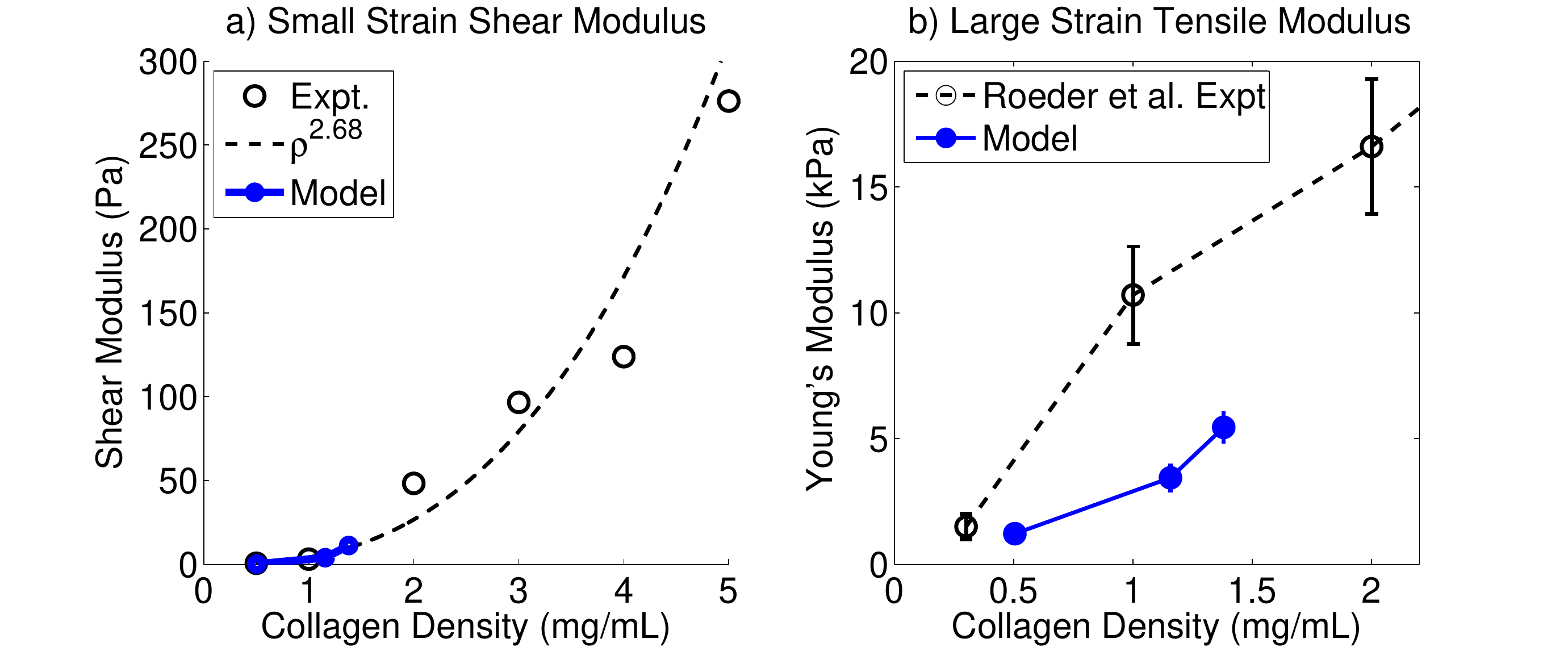}
		\caption{a) The small strain shear modulus is compared to the cone-plate rheometer experiments and a scaling law of $G' \sim \rho^{2.68}$ is observed for the experiment.  b) The large strain tensile modulus from the model and from the experiments of  Roeder et al. \cite{Roeder02}.  Results differ by a factor of 2.5, which is reasonable since the two experimental protocols were different \cite{Roeder02}.  \label{fig:GE}}
\end{figure}

The minimum energy state of the network at an imposed strain $U(\epsilon_{ij})$ is found using a conjugate gradient method developed by Hager and Zhang \cite{Hager05}.  The stress required to hold the network in its current configuration is given by $\sigma_{ij}(\epsilon_{ij}) = (dU/d \epsilon_{ij})/A$, where $A$ denotes the area of $F_0$.  The results shown in Figures~\ref{fig:strainsweep} and \ref{fig:GE} are averaged over all four extracted networks and over all 6 principle shear deformations in the sheared network and all 3 principle tensile directions in the stretched network.

The results from our shear cell experiments are given in Figures~\ref{fig:strainsweep} and \ref{fig:GE}a.  In addition, in Figure~\ref{fig:GE}b, we also present the previously reported tensile modulus of large samples that are centimeters in length \cite{Roeder02}.  Figure~\ref{fig:strainsweep}a shows a strain sweep from 0.5\% to 100\% shear strain in both the model and cone-plate rheometer experiments.  At small strains, the stress-strain response is linear, as expected, and at larger shear strains, the stress-strain response appears cubic.  In Figure~\ref{fig:GE}a, we show that the small strain modulus scales by $\sigma_{12} \sim \rho^{2.68} \epsilon_{12}$, where $\rho$ is the collagen density.  At this time, it is not possible to verify the power law scaling in the model since only densities of 0.5, 1.2, and 1.4 were observed.  The fluorescent labeling of the network changes the polymerization properties of the network, causing it to clump at higher densities.  We use this scaling relationship to collapse the curves in Figure~\ref{fig:strainsweep}b.  The close agreement between model and experiment indicates that strain stiffening due to the geometric rearrangement of the collagen fibers is enough to explain the strain stiffening seen in experiments.   

At large strains ($\sim 50\%$ in tension, $\sim 200\%$ in shear), the stress-strain curve of the model becomes linear again, though with a much steeper slope.  In Figure~\ref{fig:GE}b, we compare the large strain tensile behavior of the model to the experiments of Roeder et al. \cite{Roeder02}.  While our model underestimates their experimental measurement by a factor of 2.5, we find this to be reasonable since the two experiments used different collagen protocols.  In particular, different buffers were used.  In Figure~\ref{fig:strainsweep}, we also explore the case where $K_x = K_b = 0$, such that we have only a network of springs connected at freely rotating pin joints.  At low strains, the network can be deformed without exerting any stress, but at strains higher than 25\%, we see that this simplification adequately describes the gel.  

A topic of investigation explored by many is the validity of the assumption that these networks deform affinely \cite{Onck05, Heussinger07_thermal}.  For brevity, we state only that the deformations are highly nonaffine in these simulations, as evidenced by a visual inspection of Figure~\ref{fig:summary}d and \ref{fig:summary}e where it is obvious that many fibers leave the volume defined by an affine deformation.  

In summary, we have presented a microstructural model of a 3d biopolymer gel using a network geometry that is based on the true network architecture.  It differs from previous work in that we use realistic network architectures that have been extracted using the FIRE algorithm.  We specifically focus on the mechanics of collagen-I networks, but emphasize that this modeling approach is generalizable to other biopolymer networks.  The model has three parameters: $\{E, r, K_x \}$.  The fiber radius and tensile modulus can be measured experimentally and the model uses realistic parameters.  The cross-link torsional spring constant  must be fit to the data and we used $K_x \sim 20 K_b/l_c$.  Fitting this single parameter gives the right strain stiffening behavior for networks at three different densities at strains that vary from 0.5\% to 50\%.   This result lends support to the hypothesis put forward by Onck et al. \cite{Onck05} that strain stiffening in polymer networks, and particularly collagen-I gels, is governed by rearrangement of the gel.   

Another finding of the model is that at strains greater than 25\%, the stiffness of the gel is governed almost entirely by stretching of the fibers.  This result is relevant for collagen because cells embedded in these gels are seen to produce deformations of this order of magnitude \cite{Pizzo05}.  In modeling large systems of this type where the strains are large, it may be sufficient to treat each fiber as a spring rather than a WLC in order to reduce the computation time.  This work also demonstrates that an understanding of the cross-link mechanics in these systems is critical to understanding their mechanical properties, as has been seen previously \cite{Wagner06}.  In much of the theoretical work that has been done on random stick networks, the cross-links are treated either as freely rotating pin joints, or welded joints of fixed angle \cite{Head03pre, Onck05, Heussinger07_thermal}.  While these are sensible simplifying assumptions in developing a theory, they are not adequate for describing actual networks.  

We note that this model has been designed to capture the short time scale ($<$ 1 hr) behavior of the network, where it behaves as an elastic solid.  Such behavior requires that the cross-links be relatively fixed \cite{Barocas95}.  This simplified model provides a starting point in the development of a more complete model of collagen gel.  Ultimately, a more sophisticated approach, such as that taken by Rodney et al. \cite{Rodney05} will be necessary to capture the full dynamic behavior of the gel, where cross-links are allowed to slip and break.
\section*{Acknowledgments}
The authors would like to thank V. Barocas, T. Stylianopoulos, E. A. Sander, E. Tuzel, H. Zhang, H. Othmer, T. Jackson, P. Smereka, R. Krasny, F. MacKintosh, A. Kabla,  R. Lee, M. Dewberry, L. Kaufman and L. Jawerth, and  for their discussions.  This work is supported by NIH Bioengineering Research Partnership grant R01 CA085139-01A2 and the Institute for Mathematics and its Applications.

\bibliographystyle{apsrev}

\begin{thebibliography}{29}
\expandafter\ifx\csname natexlab\endcsname\relax\def\natexlab#1{#1}\fi
\expandafter\ifx\csname bibnamefont\endcsname\relax
  \def\bibnamefont#1{#1}\fi
\expandafter\ifx\csname bibfnamefont\endcsname\relax
  \def\bibfnamefont#1{#1}\fi
\expandafter\ifx\csname citenamefont\endcsname\relax
  \def\citenamefont#1{#1}\fi
\expandafter\ifx\csname url\endcsname\relax
  \def\url#1{\texttt{#1}}\fi
\expandafter\ifx\csname urlprefix\endcsname\relax\def\urlprefix{URL }\fi
\providecommand{\bibinfo}[2]{#2}
\providecommand{\eprint}[2][]{\url{#2}}

\bibitem[{\citenamefont{Lodish et~al.}(1999)\citenamefont{Lodish, Berk,
  Zipursky, Matsudaira, Baltimore, and Darnell}}]{Lodish99}
\bibinfo{author}{\bibfnamefont{H.}~\bibnamefont{Lodish}},
  \bibinfo{author}{\bibfnamefont{A.}~\bibnamefont{Berk}},
  \bibinfo{author}{\bibfnamefont{S.~L.} \bibnamefont{Zipursky}},
  \bibinfo{author}{\bibfnamefont{P.}~\bibnamefont{Matsudaira}},
  \bibinfo{author}{\bibfnamefont{D.}~\bibnamefont{Baltimore}},
  \bibnamefont{and} \bibinfo{author}{\bibfnamefont{J.}~\bibnamefont{Darnell}},
  \emph{\bibinfo{title}{Molecular Cell Biology}} (\bibinfo{publisher}{W. H.
  Freeman and Company}, \bibinfo{year}{1999}).

\bibitem[{\citenamefont{Fung}(1993)}]{Fung93}
\bibinfo{author}{\bibfnamefont{Y.~C.} \bibnamefont{Fung}},
  \emph{\bibinfo{title}{Biomechanics: Mechanical Properties of Living Tissues}}
  (\bibinfo{publisher}{Springer}, \bibinfo{address}{New York},
  \bibinfo{year}{1993}), \bibinfo{edition}{2nd} ed.

\bibitem[{\citenamefont{Stein et~al.}(2008)\citenamefont{Stein, Vader, Jawerth,
  Weitz, and Sander}}]{Stein08_fire}
\bibinfo{author}{\bibfnamefont{A.~M.} \bibnamefont{Stein}},
  \bibinfo{author}{\bibfnamefont{D.~A.} \bibnamefont{Vader}},
  \bibinfo{author}{\bibfnamefont{L.~M.} \bibnamefont{Jawerth}},
  \bibinfo{author}{\bibfnamefont{D.~A.} \bibnamefont{Weitz}}, \bibnamefont{and}
  \bibinfo{author}{\bibfnamefont{L.~M.} \bibnamefont{Sander}},
  \bibinfo{journal}{accepted by J. Microscopy}  (\bibinfo{year}{2008}).

\bibitem[{\citenamefont{Chandran and Barocas}(2004)}]{Chandran04}
\bibinfo{author}{\bibfnamefont{P.~L.} \bibnamefont{Chandran}} \bibnamefont{and}
  \bibinfo{author}{\bibfnamefont{V.~H.} \bibnamefont{Barocas}},
  \bibinfo{journal}{Journal of Biomechanical Engineering}
  \textbf{\bibinfo{volume}{126}}, \bibinfo{pages}{152} (\bibinfo{year}{2004}).

\bibitem[{\citenamefont{Friedl et~al.}(1998)\citenamefont{Friedl, Zanker, and
  Brocker}}]{Friedl98}
\bibinfo{author}{\bibfnamefont{P.}~\bibnamefont{Friedl}},
  \bibinfo{author}{\bibfnamefont{K.~S.} \bibnamefont{Zanker}},
  \bibnamefont{and} \bibinfo{author}{\bibfnamefont{E.-B.}
  \bibnamefont{Brocker}}, \bibinfo{journal}{Microscopy Res. and Tech.}
  \textbf{\bibinfo{volume}{43}}, \bibinfo{pages}{369} (\bibinfo{year}{1998}).

\bibitem[{\citenamefont{Kaufman et~al.}(2005)\citenamefont{Kaufman, Brangwynne,
  Kasza, Filippidi, Gordon, Deisboeck, and Weitz}}]{Kaufman05}
\bibinfo{author}{\bibfnamefont{L.~J.} \bibnamefont{Kaufman}},
  \bibinfo{author}{\bibfnamefont{C.~P.} \bibnamefont{Brangwynne}},
  \bibinfo{author}{\bibfnamefont{K.~E.} \bibnamefont{Kasza}},
  \bibinfo{author}{\bibfnamefont{E.}~\bibnamefont{Filippidi}},
  \bibinfo{author}{\bibfnamefont{V.~D.} \bibnamefont{Gordon}},
  \bibinfo{author}{\bibfnamefont{T.~S.} \bibnamefont{Deisboeck}},
  \bibnamefont{and} \bibinfo{author}{\bibfnamefont{D.~A.} \bibnamefont{Weitz}},
  \bibinfo{journal}{Biophys. J.} \textbf{\bibinfo{volume}{89}},
  \bibinfo{pages}{635} (\bibinfo{year}{2005}).

\bibitem[{\citenamefont{Stein et~al.}(2007)\citenamefont{Stein, Demuth, Mobley,
  Berens, and Sander}}]{Stein07_bj}
\bibinfo{author}{\bibfnamefont{A.~M.} \bibnamefont{Stein}},
  \bibinfo{author}{\bibfnamefont{T.}~\bibnamefont{Demuth}},
  \bibinfo{author}{\bibfnamefont{D.}~\bibnamefont{Mobley}},
  \bibinfo{author}{\bibfnamefont{M.~E.} \bibnamefont{Berens}},
  \bibnamefont{and} \bibinfo{author}{\bibfnamefont{L.~M.}
  \bibnamefont{Sander}}, \bibinfo{journal}{Biophys. J.}
  \textbf{\bibinfo{volume}{92}}, \bibinfo{pages}{356} (\bibinfo{year}{2007}).

\bibitem[{\citenamefont{Storm et~al.}(2005)\citenamefont{Storm, Patsore,
  MacKintosh, Lubensky, and Janmey}}]{Storm05}
\bibinfo{author}{\bibfnamefont{C.}~\bibnamefont{Storm}},
  \bibinfo{author}{\bibfnamefont{J.}~\bibnamefont{Patsore}},
  \bibinfo{author}{\bibfnamefont{F.}~\bibnamefont{MacKintosh}},
  \bibinfo{author}{\bibfnamefont{T.}~\bibnamefont{Lubensky}}, \bibnamefont{and}
  \bibinfo{author}{\bibfnamefont{P.}~\bibnamefont{Janmey}},
  \bibinfo{journal}{Nature} \textbf{\bibinfo{volume}{435}},
  \bibinfo{pages}{191} (\bibinfo{year}{2005}).

\bibitem[{\citenamefont{Heussinger and Frey}(2006)}]{Heussinger06}
\bibinfo{author}{\bibfnamefont{C.}~\bibnamefont{Heussinger}} \bibnamefont{and}
  \bibinfo{author}{\bibfnamefont{E.}~\bibnamefont{Frey}},
  \bibinfo{journal}{Phys. Rev. Lett.} \textbf{\bibinfo{volume}{96}},
  \bibinfo{pages}{017802} (\bibinfo{year}{2006}).

\bibitem[{\citenamefont{Heussinger et~al.}(2007)\citenamefont{Heussinger,
  Schaefer, and Frey}}]{Heussinger07}
\bibinfo{author}{\bibfnamefont{C.}~\bibnamefont{Heussinger}},
  \bibinfo{author}{\bibfnamefont{B.}~\bibnamefont{Schaefer}}, \bibnamefont{and}
  \bibinfo{author}{\bibfnamefont{E.}~\bibnamefont{Frey}},
  \bibinfo{journal}{Phys. Rev. E} \textbf{\bibinfo{volume}{76}},
  \bibinfo{pages}{031906} (\bibinfo{year}{2007}).

\bibitem[{\citenamefont{Onck et~al.}(2005)\citenamefont{Onck, Koeman, van
  Dillen, and van~der Giessen}}]{Onck05}
\bibinfo{author}{\bibfnamefont{P.}~\bibnamefont{Onck}},
  \bibinfo{author}{\bibfnamefont{T.}~\bibnamefont{Koeman}},
  \bibinfo{author}{\bibfnamefont{T.}~\bibnamefont{van Dillen}},
  \bibnamefont{and} \bibinfo{author}{\bibfnamefont{E.}~\bibnamefont{van~der
  Giessen}}, \bibinfo{journal}{Phys. Rev. Lett.} \textbf{\bibinfo{volume}{95}},
  \bibinfo{pages}{178102} (\bibinfo{year}{2005}).

\bibitem[{\citenamefont{Huisman et~al.}(2007)\citenamefont{Huisman, van Dillen,
  Onck, and van~der Giessen}}]{Huisman07}
\bibinfo{author}{\bibfnamefont{E.~M.} \bibnamefont{Huisman}},
  \bibinfo{author}{\bibfnamefont{T.}~\bibnamefont{van Dillen}},
  \bibinfo{author}{\bibfnamefont{P.}~\bibnamefont{Onck}}, \bibnamefont{and}
  \bibinfo{author}{\bibfnamefont{E.}~\bibnamefont{van~der Giessen}},
  \bibinfo{journal}{Phys. Rev. Lett.} \textbf{\bibinfo{volume}{99}},
  \bibinfo{pages}{208103} (\bibinfo{year}{2007}).

\bibitem[{\citenamefont{Chandran and Barocas}(2006)}]{Chandran06}
\bibinfo{author}{\bibfnamefont{P.~L.} \bibnamefont{Chandran}} \bibnamefont{and}
  \bibinfo{author}{\bibfnamefont{V.~H.} \bibnamefont{Barocas}},
  \bibinfo{journal}{J. Biomech. Eng.} \textbf{\bibinfo{volume}{128}},
  \bibinfo{pages}{259} (\bibinfo{year}{2006}).

\bibitem[{\citenamefont{Stylianopoulos and Barocas}(2007)}]{Stylianopoulos07}
\bibinfo{author}{\bibfnamefont{T.}~\bibnamefont{Stylianopoulos}}
  \bibnamefont{and} \bibinfo{author}{\bibfnamefont{V.~H.}
  \bibnamefont{Barocas}}, \bibinfo{journal}{Comput. Methods Appl. Mech. Eng.}
  \textbf{\bibinfo{volume}{196}}, \bibinfo{pages}{2981} (\bibinfo{year}{2007}).

\bibitem[{\citenamefont{Klapper and Qian}(1998)}]{Klapper98}
\bibinfo{author}{\bibfnamefont{I.}~\bibnamefont{Klapper}} \bibnamefont{and}
  \bibinfo{author}{\bibfnamefont{H.}~\bibnamefont{Qian}},
  \bibinfo{journal}{Biophys. J.} \textbf{\bibinfo{volume}{74}},
  \bibinfo{pages}{2504} (\bibinfo{year}{1998}).

\bibitem[{\citenamefont{Graham et~al.}(2006)\citenamefont{Graham, Vomund,
  Phillips, and Grandbois}}]{Graham06}
\bibinfo{author}{\bibfnamefont{J.~S.} \bibnamefont{Graham}},
  \bibinfo{author}{\bibfnamefont{A.~N.} \bibnamefont{Vomund}},
  \bibinfo{author}{\bibfnamefont{C.~L.} \bibnamefont{Phillips}},
  \bibnamefont{and}
  \bibinfo{author}{\bibfnamefont{M.}~\bibnamefont{Grandbois}},
  \bibinfo{journal}{Exp. Cell Res.} \textbf{\bibinfo{volume}{299}},
  \bibinfo{pages}{335} (\bibinfo{year}{2006}).

\bibitem[{\citenamefont{Miyazaki and Hayashi}(1999)}]{Miyazaki99}
\bibinfo{author}{\bibfnamefont{H.}~\bibnamefont{Miyazaki}} \bibnamefont{and}
  \bibinfo{author}{\bibfnamefont{K.}~\bibnamefont{Hayashi}},
  \bibinfo{journal}{Biomedical Microdevices} \textbf{\bibinfo{volume}{2}},
  \bibinfo{pages}{151} (\bibinfo{year}{1999}).

\bibitem[{\citenamefont{van~der Rijt et~al.}(2006)\citenamefont{van~der Rijt,
  van~der Werf, Bennink, Dijkstra, and Feijen}}]{vanderRijt06}
\bibinfo{author}{\bibfnamefont{J.~A.~J.} \bibnamefont{van~der Rijt}},
  \bibinfo{author}{\bibfnamefont{K.~O.} \bibnamefont{van~der Werf}},
  \bibinfo{author}{\bibfnamefont{M.~L.} \bibnamefont{Bennink}},
  \bibinfo{author}{\bibfnamefont{P.~J.} \bibnamefont{Dijkstra}},
  \bibnamefont{and} \bibinfo{author}{\bibfnamefont{J.}~\bibnamefont{Feijen}},
  \bibinfo{journal}{Macromol. Biosci.} \textbf{\bibinfo{volume}{6}},
  \bibinfo{pages}{697} (\bibinfo{year}{2006}).

\bibitem[{\citenamefont{Raub et~al.}(2007)\citenamefont{Raub, Suresh, Krasieva,
  Lyubovitsky, Mih, Putnam, Tromberg, and George}}]{Raub07}
\bibinfo{author}{\bibfnamefont{C.~B.} \bibnamefont{Raub}},
  \bibinfo{author}{\bibfnamefont{V.}~\bibnamefont{Suresh}},
  \bibinfo{author}{\bibfnamefont{T.}~\bibnamefont{Krasieva}},
  \bibinfo{author}{\bibfnamefont{J.}~\bibnamefont{Lyubovitsky}},
  \bibinfo{author}{\bibfnamefont{J.~D.} \bibnamefont{Mih}},
  \bibinfo{author}{\bibfnamefont{A.~J.} \bibnamefont{Putnam}},
  \bibinfo{author}{\bibfnamefont{B.~J.} \bibnamefont{Tromberg}},
  \bibnamefont{and} \bibinfo{author}{\bibfnamefont{S.~C.}
  \bibnamefont{George}}, \bibinfo{journal}{Biophys. J.}
  \textbf{\bibinfo{volume}{92}}, \bibinfo{pages}{2212} (\bibinfo{year}{2007}).

\bibitem[{\citenamefont{Yang et~al.}(2008)\citenamefont{Yang, van~der Werf,
  Koopman, Subramaniam, Bennink, Dijkstra, and Feijen}}]{Yang08}
\bibinfo{author}{\bibfnamefont{L.}~\bibnamefont{Yang}},
  \bibinfo{author}{\bibfnamefont{K.~O.} \bibnamefont{van~der Werf}},
  \bibinfo{author}{\bibfnamefont{B.~F.} \bibnamefont{Koopman}},
  \bibinfo{author}{\bibfnamefont{V.}~\bibnamefont{Subramaniam}},
  \bibinfo{author}{\bibfnamefont{M.~L.} \bibnamefont{Bennink}},
  \bibinfo{author}{\bibfnamefont{P.~J.} \bibnamefont{Dijkstra}},
  \bibnamefont{and} \bibinfo{author}{\bibfnamefont{J.}~\bibnamefont{Feijen}},
  \bibinfo{journal}{Biophys. J.} \textbf{\bibinfo{volume}{94}},
  \bibinfo{pages}{2204} (\bibinfo{year}{2008}).

\bibitem[{\citenamefont{Head et~al.}(2003)\citenamefont{Head, Levine, and
  MacKintosh}}]{Head03pre}
\bibinfo{author}{\bibfnamefont{D.}~\bibnamefont{Head}},
  \bibinfo{author}{\bibfnamefont{A.~J.} \bibnamefont{Levine}},
  \bibnamefont{and}
  \bibinfo{author}{\bibfnamefont{F.}~\bibnamefont{MacKintosh}},
  \bibinfo{journal}{Phys. Rev. E} \textbf{\bibinfo{volume}{68}},
  \bibinfo{pages}{061907} (\bibinfo{year}{2003}).

\bibitem[{\citenamefont{Roeder et~al.}(2002)\citenamefont{Roeder, Kokini,
  Sturgis, Robinson, and Voytik-Harbin}}]{Roeder02}
\bibinfo{author}{\bibfnamefont{B.~A.} \bibnamefont{Roeder}},
  \bibinfo{author}{\bibfnamefont{K.}~\bibnamefont{Kokini}},
  \bibinfo{author}{\bibfnamefont{J.~E.} \bibnamefont{Sturgis}},
  \bibinfo{author}{\bibfnamefont{J.~P.} \bibnamefont{Robinson}},
  \bibnamefont{and} \bibinfo{author}{\bibfnamefont{S.~L.}
  \bibnamefont{Voytik-Harbin}}, \bibinfo{journal}{Journal of Biomechanical
  Engineering} \textbf{\bibinfo{volume}{124}}, \bibinfo{pages}{214}
  (\bibinfo{year}{2002}).

\bibitem[{\citenamefont{Krishnan et~al.}(2004)\citenamefont{Krishnan, Weiss,
  Wessman, and Hoying}}]{Krishnan04}
\bibinfo{author}{\bibfnamefont{L.}~\bibnamefont{Krishnan}},
  \bibinfo{author}{\bibfnamefont{J.~A.} \bibnamefont{Weiss}},
  \bibinfo{author}{\bibfnamefont{M.~D.} \bibnamefont{Wessman}},
  \bibnamefont{and} \bibinfo{author}{\bibfnamefont{J.~B.}
  \bibnamefont{Hoying}}, \bibinfo{journal}{Tissue Engineering}
  \textbf{\bibinfo{volume}{10}}, \bibinfo{pages}{241} (\bibinfo{year}{2004}).

\bibitem[{\citenamefont{Hager and Zhang}(2005)}]{Hager05}
\bibinfo{author}{\bibfnamefont{W.~W.} \bibnamefont{Hager}} \bibnamefont{and}
  \bibinfo{author}{\bibfnamefont{H.}~\bibnamefont{Zhang}},
  \bibinfo{journal}{SIAM J. Optim.} \textbf{\bibinfo{volume}{16}},
  \bibinfo{pages}{170} (\bibinfo{year}{2005}).

\bibitem[{\citenamefont{Heussinger and Frey}(2007)}]{Heussinger07_thermal}
\bibinfo{author}{\bibfnamefont{C.}~\bibnamefont{Heussinger}} \bibnamefont{and}
  \bibinfo{author}{\bibfnamefont{E.}~\bibnamefont{Frey}},
  \bibinfo{journal}{Phys. Rev. E} \textbf{\bibinfo{volume}{75}},
  \bibinfo{pages}{011917} (\bibinfo{year}{2007}).

\bibitem[{\citenamefont{Pizzo et~al.}(2005)\citenamefont{Pizzo, Kokini, Vaughn,
  Waisner, and Voytik-Harbin}}]{Pizzo05}
\bibinfo{author}{\bibfnamefont{A.~M.} \bibnamefont{Pizzo}},
  \bibinfo{author}{\bibfnamefont{K.}~\bibnamefont{Kokini}},
  \bibinfo{author}{\bibfnamefont{L.~C.} \bibnamefont{Vaughn}},
  \bibinfo{author}{\bibfnamefont{B.~Z.} \bibnamefont{Waisner}},
  \bibnamefont{and} \bibinfo{author}{\bibfnamefont{S.~L.}
  \bibnamefont{Voytik-Harbin}}, \bibinfo{journal}{J. Appl. Physiol.}
  \textbf{\bibinfo{volume}{98}}, \bibinfo{pages}{1909} (\bibinfo{year}{2005}).

\bibitem[{\citenamefont{Wagner et~al.}(2006)\citenamefont{Wagner, Tharmann,
  Haase, Fischer, and Bausch}}]{Wagner06}
\bibinfo{author}{\bibfnamefont{B.}~\bibnamefont{Wagner}},
  \bibinfo{author}{\bibfnamefont{R.}~\bibnamefont{Tharmann}},
  \bibinfo{author}{\bibfnamefont{I.}~\bibnamefont{Haase}},
  \bibinfo{author}{\bibfnamefont{M.}~\bibnamefont{Fischer}}, \bibnamefont{and}
  \bibinfo{author}{\bibfnamefont{A.~R.} \bibnamefont{Bausch}},
  \bibinfo{journal}{Proc. Natl. Acad. Sci. USA} \textbf{\bibinfo{volume}{103}},
  \bibinfo{pages}{13974} (\bibinfo{year}{2006}).

\bibitem[{\citenamefont{Barocas et~al.}(1995)\citenamefont{Barocas, Moon, and
  Tranquillo}}]{Barocas95}
\bibinfo{author}{\bibfnamefont{V.~H.} \bibnamefont{Barocas}},
  \bibinfo{author}{\bibfnamefont{A.~G.} \bibnamefont{Moon}}, \bibnamefont{and}
  \bibinfo{author}{\bibfnamefont{R.~T.} \bibnamefont{Tranquillo}},
  \bibinfo{journal}{J. Biomech. Eng.} \textbf{\bibinfo{volume}{117}},
  \bibinfo{pages}{161} (\bibinfo{year}{1995}).

\bibitem[{\citenamefont{Rodney et~al.}(2005)\citenamefont{Rodney, Fivel, and
  Dendievel}}]{Rodney05}
\bibinfo{author}{\bibfnamefont{D.}~\bibnamefont{Rodney}},
  \bibinfo{author}{\bibfnamefont{M.}~\bibnamefont{Fivel}}, \bibnamefont{and}
  \bibinfo{author}{\bibfnamefont{R.}~\bibnamefont{Dendievel}},
  \bibinfo{journal}{Phys. Rev. Lett.} \textbf{\bibinfo{volume}{95}},
  \bibinfo{pages}{108004} (\bibinfo{year}{2005}).

\end{thebibliography}

\end{document}